\documentclass[12pt]{article} 

\pdfoutput=1

\usepackage[T1]{fontenc}
\usepackage[utf8]{inputenc}
\usepackage{newtxtext}
\usepackage{newtxmath}
\usepackage{microtype}

\usepackage{booktabs}

\usepackage{amsmath}

\usepackage{natbib}
\usepackage{hyperref}

\usepackage{geometry}

\makeatletter
\newif\if@anonymize
\@anonymizefalse

\newcommand{\blinded}[2]{%
\if@anonymize
#2%
\else
#1%
\fi}

\title{Expanding the scope of statistical computing: Training statisticians to
  be software engineers}
\author{Alex Reinhart and Christopher R. Genovese}
\date{\today}

\begin{document}
\maketitle

\begin{abstract}
  Traditionally, statistical computing courses have taught the syntax of a
  particular programming language or specific statistical computation methods.
  Since the publication of \citet{Nolan:2010}, we have seen a greater emphasis
  on data wrangling, reproducible research, and visualization. This shift
  better prepares students for careers working with complex datasets and
  producing analyses for multiple audiences. But, we argue, statisticians are
  now often called upon to develop statistical \emph{software}, not just
  analyses, such as R packages implementing new analysis methods or machine
  learning systems integrated into commercial products. This demands different
  skills.

  We describe a graduate course that we developed to meet this need by focusing on four
  themes: programming practices; software design; important algorithms and data
  structures; and essential tools and methods. Through code review and revision,
  and a semester-long software project, students practice all the skills of
  software engineering. The course allows students to expand their understanding
  of computing \emph{as applied to statistical problems} while building
  expertise in the kind of software development that is increasingly the
  province of the working statistician. We see this as a model for the future
  evolution of the computing curriculum in statistics and data science.
\end{abstract}


\section{Introduction}

When \citet{Nolan:2010} wrote their seminal paper on the role of computing in
statistics and statistics curricula, they noted the rapid change in the skills
needed by practicing statisticians. It would no longer be sufficient for
statisticians to learn computing only as a collection of numerical methods or
specialized statistical algorithms, such as Markov chain Monte Carlo or 
generating pseudo-random numbers. Statisticians now face large quantities of data,
often in new forms like text or networks, and this data must be obtained---such
as from Web services or databases---then managed, wrangled in complex ways, and
visualized. They argued that arming students with a solid computational base
will prepare them to adapt to the wide range of problems they will see on the
job---and that these computational skills will also give them new ways to
explore and understand the statistical concepts. They suggested syllabi and
curricula that would advance understanding of these skills in both undergraduate
and graduate programs \citep{Nolan:2009}.

This premise has only become more true in the intervening years. As the
conversation shifts to ``data science'' and organizations apply statistical
thinking to an ever wider range of problems, statisticians must use their
computational skills to acquire data from disparate sources, integrate it into a
useful form, conduct exploratory analyses and visualizations to understand the
data's full complexity, and only then use statistical procedures to draw
conclusions. To ensure these conclusions are reproducible, statisticians must
also use computational tools like knitr \citep{Xie:2015} and the command line to
automate a pipeline of scripts, analyses and results.

In this paper, we argue that though these computational skills are important,
for some statisticians they are only a fraction of what is now needed. Many
statisticians now find themselves delivering not \emph{analyses}---in the form
of reports or presentations on some statistical analysis---but \emph{products}
that are used continually. In academia, these products might be R packages
implementing a newly developed statistical method, so that others can apply the
same method to their own problems. In industry, these products could be new
methods to detect fraud or improve advertising in a large online service, used
continuously as new data arrives and new decisions have to be made. In either
case, the product is often a large and complex piece of \emph{software} with a
codebase developed by a team over many months, and it's never truly ``done'': it
must be maintained and updated as conditions change and new requirements are
placed on it.

To build and maintain these products, statisticians need additional skills.
Structuring a large and complex codebase so it can be easily understood requires
principles from software engineering; writing code with a team requires version
control systems and collaboration skills; applying new statistical methods to
large and complicated data requires a firm understanding of algorithms and data
structures so the resulting code will be efficient. And everything must be
well-tested and debugged so colleagues, bosses, and users can have confidence in
the results. These skills are less important for a one-off data analysis, but
they are crucial for the tasks statisticians face as they put their expertise
into practice as part of long-running systems and widely used products.

Beginning in 2015, we have developed a graduate-level course in statistical
computing intended to teach software engineering skills. The course is now part
of the required curriculum for both the Master's in \blinded{Statistical
  Practice}{Statistics} and the Ph.D.\ in Statistics \& Data Science at
\blinded{Carnegie Mellon University}{[our institution]}, serving roughly 40--50
students per year. Most of these students have prior statistical programming
experience from undergraduate courses, and our course is their only required
graduate-level statistical computing course. The students have widely varied
backgrounds, and though the master's program emphasizes professional skills
\citep[see][]{Greenhouse:2018} while the Ph.D.\ program emphasizes theoretical
and applied research, the course goals are shared: to prepare students to build
complex statistical software.

In this paper, we set out the skills we aim to teach and the strategies we use
to teach them. Our pedagogy has evolved every year as we have discovered that
the course pedagogy is inextricable from its content. For students to learn
complex computational skills, we must give them regular practice with these
skills and rapid, targeted feedback on their performance. We argue that these skills are
becoming increasingly important for graduate-level statisticians and cannot be
left to others to fill in. Computation has only grown in importance in the ten
years since Nolan and Temple Lang issued their call to action, and we expect it
will only become \emph{more} important in the ten years to come.

\section{Role of Computing in Statistics and Data Science}

Before we discuss the statistical computing course we developed, it will be
useful to briefly trace the evolution of computing's role in statistics and
data science.

Since roughly 2000, a major focus of work and teaching in statistical computing
has been ``Literate Statistical Practice'', which ``encourages the construction
of documentation for data management and statistical analysis as the code for it
is produced'' \citep{Rossini:2001}. Tools such as Sweave \citep{Leisch:2002}
began to allow statisticians to embed the code producing their analysis
\emph{inside} the text of the analysis report, so that a single command would
run the analysis, produce the results, format tables and figures, and typeset
the report for distribution. This had many practical advantages, making it easy
to make small changes and then re-run analyses and reports from scratch, and has
become even more important as reproducibility has become a major concern. Tools
like knitr and R Markdown \citep{Xie:2015,Xie:2018} have made reproducible
reports easier to write and easier to use, contributing to their rapid spread.

These tools are now widely used in statistics education and in practice.
\citet{Baumer:2014wi}, for example, use knitr in introductory statistics courses
to ``develop the basic capacity to undertake modern data analysis and
communicate their results,'' and \citet{CetinkayaRundel:2018} described its
coordinated use in an undergraduate course sequence designed to develop
statistical computing skills in students from the introductory level onward. In
industry, \citet{Bion:2018fn} describe the widespread adoption of knitr among
the data science team at Airbnb, who use it routinely to share their analyses
and business experiments with each other, with management, and even publicly as
blog posts and academic publications.

The other main emphasis of statistical computing work has been on tools to make
wrangling, restructuring, summarizing, and aggregating data much easier. There
is a growing emphasis on ``tidy data'' \citep{Wickham_2014}, and the R community
has developed many new packages \citep[e.g.\ the Tidyverse,][]{Wickham_2019}
that make it easy for statisticians to express the operations they need to
wrangle their data into the most convenient form. Other packages facilitate
interactive visualizations or make it easy to present statistical results in
written reports. Statistical computing curricula have adapted to include these
tools and to give students authentic practice wrangling messy data.

But we should not narrow our focus too quickly. Not every statistical task fits
into the framework of ``receive a question, wrangle the data, conduct an
analysis, and write a report on the results.'' No longer relegated to roles as
consultants or analysts brought in to answer specific questions, statisticians
and data scientists increasingly hold roles as integral parts of teams
developing products and delivering services. They need ``data acumen,'' which
includes facility with a much wider range of tools and the ability to
collaborate with software engineering teams and other disciplines \citep[Chapter
2]{NAS:2018}.

\citet{Bion:2018fn} provided an insightful example of this shift. At Airbnb, a
service allowing property owners to list spaces for short-term rentals, the data
science team might build ``a machine learning algorithm that takes into account
a variety of points of information'' to suggest a fair price for a host to
charge for guests. But the outcome of this work is not a report to be submitted
to management describing the results of their modeling efforts---after
developing a prototype model, they ``worked with engineers to bring the
prototype into production,'' where hosts now use its recommendations every day.
That is, the final outcome was the \emph{deployment} of a piece of statistical
software, which now continually operates as a part of Airbnb's core business.

We can also look beyond industry to see statistical software used for purposes
other than writing analytical reports. Consider a Ph.D.\ student conducting
theoretical work on new models for some complex type of data. This work may
involve thousands of lines of code: code to simulate data with known parameters,
code to estimate the model from data, code to run simulation studies that verify
theoretical results, code to calculate diagnostics or measure goodness of fit,
code to fit benchmark models and run comparisons, \textit{etc}. Much of the code
forms a \emph{product} that an ambitious graduate student might release as an R
package submitted to CRAN or a Python package on PyPI, allowing other
researchers to benefit from their theoretical labor and use the newly developed
methods for their own practical purposes. And the wide availability of these
statistical products has been a boon for the field, allowing new statistical
methods to be quickly adopted in industry \citep{Bion:2018fn}. The broader
impact of this statistical software ecosystem is hard to overstate.

The shift in statistical computing is noticeable in the work done by
statisticians in academia, but also by the jobs they take in industry. For
example, of 91 total graduates of \blinded{Carnegie Mellon University's}{our
  institution's} Bachelor's in Statistics \& Data Science program in 2018, 56
reported being employed at the time of a survey of their career outcomes, and of
these, 14 (25\%) reported a job title implying a software development role, such
as ``Software Engineer'' \blinded{\citep{CPDC:2018}}{[citation blinded]}. It has
  been our experience that many industry roles titled ``Data Scientist'' or
  ``Data Analyst'' also heavily involve software development.

\section{Course Content}

In the next sections, we discuss what it would mean for a statistical computing
curriculum to prepare students for these roles, and discuss a course we developed
to do so. We focus on four themes---four sets of skills students must learn to
effectively develop statistical products and not just statistical reports.
These themes are covered in lectures, but it is also vital that the course give
students repeated practice with all these skills, and the necessary assignments
and pedagogy will be discussed in Section~\ref{pedagogy}.

\subsection{Four Themes for Statistical Product Development}
\label{four-themes}

\subsubsection{Effective Programming Practices}
\label{programming-practices}

Students must learn practices that make software more reliable, more usable, and
easier to maintain. Such practices include testing, code review, clear naming,
and effective documentation.

Unit testing, for example, is often adopted in professional software development
to ensure code is correct and defects are not accidentally introduced. A unit
test isolates a specific ``unit'' of code, such as a function or class, and runs
that unit with specific inputs, then verifies that the unit's output matches the
expected output. Unit tests are written using a package designed to organize
test cases, run all tests automatically with a single command, and report
summary results indicating which test cases failed and giving descriptions of
the failure. The \texttt{testthat} package is widely used for R
\citep{Wickham:2011}, and similar packages are available for almost every common
programming language.

Unit testing is an essential part of software engineering for several reasons.
Most obviously, it helps ensure correctness of software. If each function or
method has detailed test cases, and these test cases can easily be checked every
time the code is changed, mistakes can be detected immediately. Software
engineering research shows that while writing unit tests takes extra time, this
time can in some settings be made up in the time saved fixing problems and
debugging errors \citep{Williams_2003,Bissi_2016}. There have been notable cases
of errors in statistical and scientific software going undetected for years,
even as the software was used routinely for scientific research, underscoring
the importance of effective testing \citep{Eklund:2016}. Less obviously,
dividing up complex tasks into simple pieces---so they can be easily
tested---also encourages software to be composed of small, easily understood
pieces, which is a key software design recommendation (see
Section~\ref{software-design}).

Code review is another essential programming practice. In collaborative software
projects, such as software developed by a team in a large company or an
open-source package developed by a group of volunteers, collaborators often
practice \emph{peer code review} \citep{Rigby:2013,Sadowski:2018}. Each proposed
change to the software, such as a new feature or fix for a problem, is submitted
for review by a coworker or collaborator. The peer gives line-by-line feedback
on the code, enforcing project style guidelines, looking for flawed reasoning
and bugs, and giving feedback so the code can be improved. Only after the
proposed change has passed peer review is it merged into the product or package.

Experiments have shown that code review detects bugs and improves software
quality, often by encouraging code to be clearer and easier to maintain
\citep{Mantyla:2009,Beller:2014}. Popular software collaboration platforms like
GitHub and GitLab support code review through ``pull requests'' or ``merge
requests.'' We give students experience with code review in two ways. We first
host an in-class activity in which students reviewed real code written (by a
course instructor) to solve a specific problem. Students are given a code review
checklist to follow, encouraging them to look at specific features of the code
and comment on them as part of their review. Later in the semester, students
conduct in-class peer code review of their Challenge projects (see
Section~\ref{challenge-project}) using GitHub's code review features.

\subsubsection{Fundamental Principles of Software Design}
\label{software-design}

Throughout the semester, we emphasize a few key principles of \emph{design}.
This includes modularity and code organization,
the way that the many features required of software are organized
into files, functions, classes, scripts, and so on.
Effective software design is a powerful means to manage complexity.
In a poor design, functions may become
large and complicated, and interact with each other in complicated ways, so that
changing one small part of the code's behavior requires intricate surgery on
many separate functions. In an effective design, functions are small and
modular, and features are clearly separated so that changing behavior only
requires changing a few specific functions that are clearly responsible for that
behavior. Good design also facilitates code reuse and refinement.

This kind of design is not a major concern when writing a literate statistical
report, which is mostly linear with a few helper functions. But when developing
a software package that's intended to be reusable, careful design is
essential---a good design makes it easy to modify and extend the package, for
example as a Ph.D.~student explores new methods in a thesis, while a bad design
can make changes excruciatingly difficult.

The semester-long Challenge project, described below in
Section~\ref{challenge-project}, is designed to give students practical design
experience. Since the project requires students to build a complicated product
over the entire semester, and later portions of the project require students to
build on or modify earlier portions, they either experience the benefits of
well-designed code or suffer the pain of modifying poorly designed code. The
teaching assistants also provide extensive feedback on design, starting before
students implement any features.

\subsubsection{Important Algorithms, Data Structures, and Representations}

In recent years, a large amount of statistical research has been focused on
\emph{scaling} statistical methods to enormous datasets without an extravagantly
large computational budget. Commonly, statistical computing courses prepare
students to work with large datasets by teaching them different tools. SQL
database systems, for example, are designed to efficiently query massive
datasets that do not fit in memory, while software like Hadoop and Apache Spark
are designed to distribute calculations across multiple servers that each have
their own chunk of data. Students might also learn to use tools like Rcpp, which
allows users to write the most computationally intensive parts of their R
packages in efficient C++ code that can be easily called from within R
\citep{Eddelbuettel:2011}. (Cython \citep{behnel2010cython} serves a similar
role in the Python world.) And students are often exposed to R programming folk
wisdom: use built-in functions whenever possible, avoid \texttt{for} loops in
favor of vectorization, and perhaps use packages like \texttt{data.table}
instead of native data frames.

But this misses the ways that careful software design can make code efficient
and scalable. First, the designer must select an \emph{algorithm} appropriate to
the task at hand, meaning the designer must be familiar with general strategies
for designing algorithms. For example, the divide-and-conquer strategy is to
reduce a large problem into several smaller problems whose solutions can be
combined to yield the overall solution; by doing this recursively, a complex
problem can be reduced to many small and trivial problems. The
divide-and-conquer strategy is widely used in computer science to produce
algorithms that scale well to large datasets (for example, mergesort is a
divide-and-conquer sorting algorithm), and it has been recently explored as a
tool for implementing statistical methods on large datasets \citep{Jordan_2013}.
Dynamic programming is another widely used strategy to break problems into
smaller problems whose solutions can be combined efficiently; for example, the
fused lasso can be expressed as a dynamic programming problem, leading to a
linear-time algorithm \citep{Johnson_2013}.

Along with the appropriate algorithm, the designer must also select appropriate
\emph{data structures} to store the data needed for an algorithm in an efficient
way. Students used to working in R for data analysis tend to think of data
frames, lists, matrices, and vectors as the only available data structures, and
often write algorithms that require repeatedly scanning through an entire
dataset to find relevant elements---which scales poorly to large datasets. But
data structures like hash tables (dictionaries), binary trees, stacks, and
queues all have their uses in statistical algorithms.

In statistics, for example, the \(k\)-d tree can store \(n\) data points, each
in \(k\) dimensions, and can find all data points in specific intervals or
ranges in \(O(\log n)\) time, rather than requiring a loop through all \(n\)
data points \citep{Bentley_1975}. This can also be used to solve
\(k\)-nearest-neighbor problems efficiently, and has been adapted to perform
fast approximate kernel density estimation \citep{Gray_2003}. Other tree data
structures are widely used by SQL databases to efficiently process queries with
complex joins and \texttt{WHERE} clauses.

Our Statistical Computing course covers basic algorithmic strategies such as
divide-and-conquer and dynamic programming, as well as basic data structures. We
emphasize to students that selecting the appropriate algorithm and data
structure can be much more important than the ordinary R performance tips. An
algorithm that uses repeated (but vectorized) scans through an entire data frame
or vector is intrinsically less efficient than one that uses a tree to do the
same operation in \(O(\log n)\) time, for example.

In the course, various homework assignments pose simple problems that can be
solved in an obvious but tremendously inefficient way as well as a less-obvious
but efficient way using an appropriate algorithm and data structure. (These can
be challenging in R, which does not provide efficient data structures by
default; for example, looking up an item by name in an R list requires an
\(O(n)\) scan through all entries, and base R does not provide flexible
collection data structures \citep{Barry_2018}.) Along with the Challenge
projects, these assignments teach students that fast code often requires careful
thinking about the organization and use of data.

\subsubsection{Essential Tools and Methods}

There are many ways to produce good statistical software, but there
are several core tools that are almost universally useful.
Such tools include editors, integrated development environments,
version control systems, debuggers and profilers,
databases (relational and otherwise), and the command line.
This theme focuses on giving students substantive experience with those
tools to give them a foundation for building good practices and habits
going forward.

We build experience with such tools into the structure of the course,
providing support for a range of quality tools while giving students
as much flexibility as possible.
For instance, we cover using SQL in class and let students
interact with SQL databases through their favorite programming language
in assignments and class activities.
Similarly,
while it is possible to work completely through graphical user interfaces (GUIs),
we believe that command line tools can add value for practioners and be a powerful tool
in many circumstances.
We show students how to use these tools and build a set of practices
for the effective design and use of such tools.

Version control is a more challenging example.
It is a critical tool for successfully developing large-scale software in collaboration with others,
allowing team members to track the history of every
source file. It allows changes to be systematically recorded and reverted if
necessary, and allows collaborators working independently to make changes to
code without interfering with each other's work. Version control software is now
widely used by companies and by collaborative
open-source software projects. The R system itself, for example, is developed
using the Subversion version control system.

There are many version control systems available, with non-trivial differences in
use and details. We focus particularly
on Git, which is perhaps the most widely used modern
version control system, particularly with the growth of web-based collaboration
services such as GitHub, GitLab, and Bitbucket that enhance Git with online tools
for filing bug reports, reviewing proposed changes to code, and tracking project
timelines and milestones. Students who are familiar with Git will be
prepared to work at organizations that use Git or similar systems, or to collaborate
on any of the thousands of open source data science packages that organize their
development with Git. \citet{Bryan:2018kq} has also persuasively argued that Git
is valuable for managing the data, code, and figures involved in a literate
statistical analysis, such as data analysis reports, further enhancing their
reproducibility by making the history of changes visible.

Unfortunately, Git is not known for being user-friendly. Its primary interface
is through the command line shell, and its documentation can be almost
impenetrable to new users. We have found that simply teaching the concepts to
students in a lecture is not sufficient; students need extensive practice
\emph{using} Git throughout the semester to begin to grasp its concepts. Hence
students use Git and GitHub to submit all homework assignments and course
projects, starting with an in-class tutorial during the first week.

Readers interested in using Git in their own courses may benefit from the
experiences of \citet{CetinkayaRundel:2018} and \citet{Fiksel:2019}, who discuss
how to use Git in statistics courses and describe common student experiences,
many of which match what we have seen in our course.

\subsection{Anti-Themes}

It would perhaps be most accurate to say that our course teaches a
problem-solving \emph{philosophy}, encompassed in the four themes presented
above, rather than simply a collection of tools suited for specific problems.
This is reflected by several topics we choose \emph{not} to cover in the course.

For example, our course \emph{does not teach a programming language}. We assume
that our students have already had some exposure to programming, such as in an
undergraduate statistical computing course or through practical experience
conducting data analyses, and so we do not spend class time covering syntax or
programming constructs. We do not require students to use any specific
programming language for their work, and examples in our lectures are often
given in R, Python, C++, Clojure, Racket, and other languages.

As the concepts, practices, and skills covered in the course are widely
relevant, this design decision makes the course accessible to students from a
range of programming backgrounds. We believe an even bigger benefit of this
approach is the perspective it offers. A focus on a single language tends to
conflate approaches to problems with the way their solutions can be expressed in
that language. Instead, we often show examples in multiple languages so that
students can see both the commonalities that are conserved across most languages
as well as some contrast across other possible design choices and idioms.
Students quickly find that, even excepting a few syntactic details, they can
understand the approach taken in a wide range of languages and that this affects
how they approach problems even in their chosen language.

We encourage students to get some experience in a new language, even if only on
simple problems. Some students use this opportunity to explore languages they
expect to use in practice (such as Python, or C++ for use with Rcpp), while
others explore more widely and pick up functional or strongly typed languages.

Similarly, the Statistical Computing course does not cover specific packages,
such as the Tidyverse \citep{Wickham_2019} or tools to obtain and wrangle data
(such as Web scraping systems). Such tools are important in practice, but
we feel it is more important for the course to cover fundamental computing
concepts that will enable students to effectively use whatever tools may appear.

\section{Pedagogy}
\label{pedagogy}

One might suspect that a computing course emphasizing concepts without
teaching any specific programming languages or tools---as our does---can't teach
the practical skills they need. But in a course intended to teach students a
complex skill, such as engineering statistical software, the only way for
students to learn the \emph{skill}---and not just the prerequisite knowledge for
that skill---is regular practice with targeted feedback. Regular practice gives
students the opportunity to practice the skills we teach, while targeted
feedback ensures they learn those skills and learn from their mistakes.

Hence the content of the Statistical Computing course cannot be separated from
its pedagogy. In this section, we describe a course design that ensures students
gain regular hands-on practice and detailed feedback, and the practical
considerations that went into it. Many of these pedagogical features were
developed through experience with each iteration of the course, and so the
course has changed significantly over time; these changes are summarized in
Table~\ref{course-rev-history} and discussed in the subsections that follow.

\begin{table}
  \centering
  \begin{tabular}{r l}\toprule
    \textbf{Semester} & \textbf{Changes} \\\midrule
    Spring 2015 & Pilot version (half-semester) \\
    Fall 2015 & Challenge project introduced (one short project) \\
    Fall 2016 & Pull request \& revision system; Master's students join; 2
                Challenge projects \\
    Fall 2017 & Various small content \& activity improvements \\
    Fall 2018 & Switch to one four-part challenge \\
    Fall 2019 & Recommended homework schedule provided \\
    Fall 2020 & Problem bank rotation; Master's students in separate course
    \\\bottomrule
  \end{tabular}
  \caption{A summary of revisions and changes made to the Statistical Computing
    course during each iteration, as discussed in Section~\ref{pedagogy}. The
    course structure has undergone many adjustments in response to experience.}
  \label{course-rev-history}
\end{table}

\subsection{Active Learning}
\label{active-learning}

In-class active learning has been repeatedly shown to improve student learning
in a variety of STEM fields \citep{Freeman_2014}. Most of our course lectures
incorporate active learning activities in various forms. For example, early in
the semester we cover unit testing; we have found that students often struggle
to think of test cases for code they write, so a large portion of the
unit-testing class is spent having the students work in small groups to think of
test cases for a few example functions.

Our experience has been that much of the student learning in a lecture seems to
come from these activities. We frequently discover that after spending 30
minutes lecturing on a particular topic and feeling that the lecture is going
well, an in-class activity reveals that some students are still deeply confused
and have misinterpreted much of the lecture. Without these activities, the
confusion could only be detected (and corrected) much later.

For key concepts that all students must master to succeed in the course, we have
gradually shifted from long lectures to in-class group activities that students
can later turn in individually for homework credit, ensuring that all students
practice the necessary skills before completing other assignments or the
Challenge project.

\subsection{Homework Problem Bank}
\label{problem-bank}

Because our students have varied levels of programming experience and have
varied goals for the course, we felt that an ordinary homework assignment
structure, where each student completes the same assignments, would be
inadequate. Some students may already be familiar with certain topics and
require little practice, while others may be most interested in a specific
topics they expect to use in their research or future job and desire specific
practice for that topic.

To allow students to choose assignments that suit their needs, we developed a
problem bank containing over 70 programming problems, categorized
by topic and difficulty level. Some problems have direct connections to
statistics, while others simply illustrate programming principles. Example
assignments include:
\begin{itemize}
\item Implement a kernel smoothing procedure, but allow the user to pass in any
  kernel function and metric of their choice, not limited by any pre-specified
  list of kernels and metrics built in to the code.
\item Implement graph search algorithms to solve mazes.
\item Implement simple text tokenization to produce bag-of-words vectors for
  documents, then explore different distance metrics between documents of
  different types.
\end{itemize}
To give students constant practice the themes discussed in
Section~\ref{four-themes}, students are required to write unit tests for every
homework assignment, and must submit them for review using Git. An automated
system runs the unit tests submitted with each homework assignment and verifies
that all tests pass.

Throughout the semester, assignments from the problem bank are posted as the
relevant topics are covered. Students can select from all the posted assignments
those they believe are most interesting or relevant to their needs, and complete
the assignments at their own pace. They may use any programming language that at
least one of the course instructors or TAs is able to read, though most students
choose Python or R. Our grading system (see Section~\ref{spec-grading}) simply
requires students to satisfactorily complete a certain number of assignments by
the end of the semester.

To encourage effective time management by students, who may be tempted to abuse
the flexibility of the homework system to put off submitting assignments until
near the end of the semester, we have used two different strategies. Our first
approach set a schedule by which students are expected to complete certain
numbers of homework points (see Section~\ref{spec-grading}); our second approach
provides a rotating selection of assignments and retires assignments from the
problem bank after 2--3 weeks, so students must complete an assignment quickly
before it becomes unavailable. This also ensures that in a given week, the TAs
must only grade a few different types of assignment, allowing them to more
efficiently grade.

\subsection{Challenge Project}
\label{challenge-project}

The homework problem bank allowed students to gain practice in many topics
covered in the course, but small homework assignments do not cover a key
learning goal of the course: learning effective strategies to design and develop
large-scale software---with all the complexity it entails---over the course of
months or years. To achieve this, the course includes a semester-long Challenge
Project. Early in the semester, students choose between several Challenges on
varying topics, and work on their chosen project for the rest of the semester.

For example, one Challenge asks students to implement an algorithm to build and
prune classification trees, then use this code to build random forest
classifiers \citep{Breiman:2001}. They then extend this code to build
classification trees using data stored in a SQL database without loading this
data into memory, in principle allowing the construction of trees for very large
datasets. Finally, they scrape abstracts from the arXiv preprint server and use
their random forest code to try to classify abstracts by subject category using
features extracted from the abstracts.

Initially the Challenge projects were designed to take several weeks and were
due in one unit. But since Fall 2018, the Challenge projects are broken into
four parts, due regularly throughout the semester. The entire project takes
roughly three months. This allows the projects to be more detailed and
ambitious, but also allows crucial scaffolding. In the first part, students
consider the design of their code but do not actually implement it. Instead,
they write function signatures but leave the bodies of the functions empty. The
only code required to be submitted is extensive unit tests demonstrating what
the code \emph{should} do, encouraging students to think more deeply about their
design before plunging in to implementation. The subsequent Challenge parts ask
students to successively add features, following the requirements given in the
assignment.

Besides the classification tree Challenge, other topics include applying the
isolation forest method for anomaly detection \citep{Liu:2012} to videos,
implementing fast data structures and algorithms for autocompletion
\citep{Wayne:2016}, and using audio fingerprinting to match short snippets of
audios to a database of recorded music \citep{Wang:2003}.
All the Challenges are designed to produce a working piece of software that
is usable in a relevant context, e.g., a package or library, a web or mobile app,
or a command-line tool.
All the Challenges are also designed to integrate multiple skills: students must select appropriate data
structures and algorithms for their code to work efficiently, while using
software design principles to keep it simple and easy to maintain.

\subsection{Specification-Based Grading and Revision}
\label{spec-grading}

During the initial iteration of the course, homework grading was fairly
conventional: the teaching assistant graded code submissions using a simple
rubric, assigning a point value to each rubric category. However, we quickly
found this system to be ineffective, as students were not reviewing the feedback
or using it to improve their future submissions.

Beginning in Fall 2016, we switched to a new system. Students select assignments
from the problem bank and submit their code through GitHub as pull requests. The
teaching assistants then give detailed line-by-line feedback on the pull
requests using GitHub's code review features. The reviews point out bugs,
critique difficult-to-read or poorly formatted code, suggest more appropriate
algorithms or data structures, request additional unit test cases, and note
design choices that make the code difficult to reuse or modify.

Crucially, there are only two possible outcomes of review: the assignment can be
marked Mastered, indicating the student has successfully solved the problem, has
used appropriate algorithms and data structures, and has written the code with
good style and with unit tests that verified its correctness; or, if those
criteria are not met, the assignment is marked ``Changes requested'' and the
student is asked to revise it according to the feedback. Once revisions are
complete, the student submits them for another review.

This system allows us to hold assignments to a very high standard. We expect
that a large fraction of submissions will be revised. (We cut the required
number of homework assignments in half at the same time as introducing the
revision system; student workload has not appreciably dropped, showing that
students are spending much more time on each assignment.) The revision process
gives students practice with a constellation of skills that are often neglected
in instruction and ensures they master the practical details of the concepts
covered in the course. One could even consider the code reviews to be
personalized tutoring provided by the teaching assistants, complementing the
lectures and activities led by the instructors. This tutoring is what allows the
course to cover topics at a high level and expect students to learn to implement
them in their chosen programming languages.

A similar revision system is used for the Challenge projects. Each part of the
Challenge is submitted to the teaching assistants for review as it is completed,
and the student must make satisfactory revisions before they can submit the next
part of the Challenge. Once all parts of the Challenge are complete and meet the
requirements, it can be graded either Mastered or Sophisticated. Sophisticated
submissions are those that demonstrate exceptional software engineering skill,
by being well-designed, clearly written, thoroughly tested, unusually
flexible and modular, and incorporating apt choices of methods/algorithms and data
organization. Earning a Sophisticated grade on the Challenge increases
the student's course grade, as discussed below in Section~\ref{grading-system}.

Prior research on mastery learning systems suggests strategies like this can
improve student learning \citep{Kulik:1990zp}, though the additional flexibility
in our system distinguishes it from the more widely used mastery grading
systems.

\subsection{Grading System}
\label{grading-system}

The course structure poses challenges for assigning final course grades. As
described in Section~\ref{problem-bank}, students select homework assignments
from a bank of possible problems. Students can complete assignments in any
order, and there are no fixed deadlines for submitting individual assignments,
nor are there points to be averaged to give a final grade.

We base grades on the number of Mastered assignments. A simple table in the
course syllabus specifies how many assignments must be Mastered to achieve a
certain grade. The Challenge project is required, but achieving a Sophisticated
on the Challenge can also move the final course grade up one grade level.

To account for the fact that individual homework assignments may involve
different degrees of difficulty, we assigned each homework assignment a certain
number of ``points.'' Typical assignments were 2 points, but difficult
assignments were 3 points and trivial assignments 1 point. The only possible
outcome is still either Mastered, meaning the student receives all the points,
or revision, meaning the student does not yet receive any points for the
assignment. There is no partial credit for submissions. The grade table is then
based on the number of points Mastered, rather than the number of assignments,
and accounts for assignment difficulty, preventing students from simply choosing
the easiest assignments to complete.

We found that this grading system has several advantages. It is noticeably
simpler than a normal points-based system to grade and administer, reducing
workload on the TAs. It reduces uncertainty for students, who know that if
they revise their submissions as instructed, they will obtain a certain number
of points, and these points translate into grades. There is no concern over a
final exam that heavily affects final grades---there is no final exam---and
students know exactly how much work they must do for a certain grade.
It also gives the students the flexibility to explore the problem bank to
improve their skills and gives them incentive to tackle some of the more
challenging problems.

\section{Conclusion}

The trends that motivated Nolan and Temple Lang's call for a new focus
on computing in statistics curricula have only accelerated.
The scope and complexity of computing tasks expected of statisticians
and data scientists require not only a detailed knowledge of statistical methods and numerical
approaches but also skills related to data management, collaboration,
and software engineering.
Our Statistical Computing course is designed to give students a firm
foundation---and authentic practice---in these skills.
It is intended to serve as a base on which their programming experience
can be built throughout their graduate career, and beyond.
Novel features of our course include emphasis on the practice of software
design, our multi-path problem bank, our grading system, integrated code
review, regular revision, and a language-agnostic approach.
The course has been successful in both our Ph.D. and Master's program.
Implementation, particularly at scale, is a continuing challenge,
and we will continue to develop and refine the course.

Statistical computing is a broad topic, and students come with varied
backgrounds and downstream needs. There are many reasonable approaches to
teaching students the computing skills they will need in their careers. We
believe, however, that working statisticians in industry and academia face
increasingly stringent demands on the capabilities, usability, and maintenance
of the software they produce, and that literate report-writing is only one
component of the many computational skills a successful statistician will need.
As the field's computing curricula continue to evolve, we believe that this
reality needs to be faced head on.

\subsection{Student Feedback}

Our university conducts anonymous course evaluation surveys at the end of each
semester; student participation is voluntary and response rates can vary widely.
Nonetheless, the quantitative data and comments from students can sometimes
provide useful information about how a course is being received.

According to the survey results, students in our statistical computing course
report working roughly 11 hours per week on the course, which is above the
intended average of 9 for a course worth its number of credits. Student comments
attributed this to the fast pace of the course: for example, one student wrote
that ``As the class was designed covers a lot of topics that would take a couple
semesters in normal C.S. courses, this course is definitely conceptually
difficult and has a quite heavy workload.'' We do not think this is an unfair
characterization, particularly for students with less prior programming
experience, and continue to adjust the curriculum and pace based on student
feedback.

Nonetheless, students enjoyed the pedagogy of the course and its flexibility.
One student noted that ``My favorite parts were the interactive parts;'' the
interactive activities discussed in Section~\ref{active-learning} ``helped me
feel more engaged and helped me understand the problems better.'' Similarly, one
student noted that ``I could pick and choose easier and harder assignments, and
get to explore new areas that interested me without being overwhelmed and
stressed out constantly.'' Though this flexibility was appreciated, it also has
its drawbacks, as noted by the student who complained that ``The homework system
also really opened my eyes to how bad I am with time management.'' With no
homework deadlines during the semester, some students experienced a mad rush to
get the required number of assignments completed in the last few weeks.

\subsection{Future Improvements}

The topics we emphasize have changed from year to year as our understanding of
student needs have changed, and the prior skills of each student cohort have
varied significantly from year to year. Also, the unconventional course
structure, while giving students significant freedom to explore their interests,
has required a great deal of experimentation to improve, and likely will
continue to change each year as we learn what structure best teaches our
intended skills.

Several challenges remain to be solved. Git has proven to be a major obstacle to
students; they must use it to submit each homework assignment on GitHub for
review by the TAs, but students who make mistakes often attempt to fix them with
ad-hoc solutions found online, leading to tangled Git histories that must be
carefully un-tangled by the instructors or TAs before assignments can be graded.
The flexible homework system can sometimes be too flexible, and without formal
deadlines, students can procrastinate and get into difficult situations, or skip
assignments selected as in-class activities and miss important skills needed for
the Challenge project. (This has been a bigger issue for Master's students
than for Ph.D. students.) The demands on course TAs to give high-quality feedback
on assignments while also holding regular office hours can be stressful,
requiring skilled TAs and large time commitments. We are working to streamline
the course, automate some aspects of homework submission and review, and improve
the student experience.

\subsection{Implementing a Similar Course}

For those interested in teaching the core themes we describe in
Section~\ref{four-themes}, comprehensive lecture notes are available at our
course website, \blinded{\url{https://36-750.github.io/}}{[blinded]}. (The
website includes more than one semester worth of material: it includes notes on
every topic that has been taught in the course, even as the selection of topics
has changed from year to year.) The notes include in-class active learning
activities, example programs, and notes that were used during lectures. The
homework problem bank (Section~\ref{problem-bank}), including solutions to some
problems, is kept privately by the authors and is available to instructors on
request.

But so far, we have left one key question open: Who can effectively teach a
statistical computing course on the topics we describe? The four core themes
require faculty with experience designing and implementing complex software;
while an introductory R programming class simply requires knowledge of syntax
and some basic principles, our themes include principles of algorithms, data
structures, and software design. To give effective feedback, the course teaching
assistants must also be experienced programmers who can recognize inefficient
algorithms or unnecessarily complex designs.

These constraints limit who can teach a course covering the skills we feel are
most important, at least until such teaching becomes more widespread and faculty
can be expected to have these skills. It may be practical, however, to co-teach
the class. An instructor experienced in statistics and data science could cover
those topics, while an instructor experienced in software engineering, perhaps
from another department, provides the core computing content.

This raises a question: Why not have students take a computer science or
software engineering course from another department?
While a fair portion of the material we emphasize (including algorithms, data structures,
testing, software design, and wide-ranging assignments) might seem more naturally obtained
from a Computer Science department, we have found many reasons to prefer
that material within a Statistics curriculum.
First, we \emph{explicitly} address these themes, with significant class time;
these skills tend to be threaded more implicitly throughout a typical computer
science curriculum.
Second, we can focus the practice of our target skills with context and examples
that are meaningful to Statistics and Data Science students.
Third, having a single foundation course early in the Statistics graduate curriculum
has been a significant downstream productivity enhancer for our students, who
soon use the skills in their other courses and projects.
Finally, we know of no other course, in Computer Science or elsewhere, that
achieves our target balance on our themes and skill development.

\if@anonymize
\else
\section*{Acknowledgments}

We thank Peter Freeman for contributions to the course design and content during
its first iteration. We are indebted to our excellent teaching assistants for
their outstanding work in the class: Philipp Burckhardt, Niccolò Dalmasso,
Sangwon Hyun, Nicolás Kim, Francis Kovacs, Taylor Pospisil, and Shamindra
Shrotriya. Jerzy Wieczorek provided helpful insight on our mastery grading
system. We thank the peer reviewers and guest editor for many suggestions that
improved the manuscript.

\fi

\bibliography{statcomp-refs}

\begin{thebibliography}{37}
\providecommand{\natexlab}[1]{#1}
\providecommand{\url}[1]{\texttt{#1}}
\expandafter\ifx\csname urlstyle\endcsname\relax
  \providecommand{\doi}[1]{doi: #1}\else
  \providecommand{\doi}{doi: \begingroup \urlstyle{rm}\Url}\fi

\bibitem[Barry(2018)]{Barry_2018}
Timothy Barry.
\newblock Collections in {R}: Review and proposal.
\newblock \emph{{The R Journal}}, 10\penalty0 (1):\penalty0 455--471, 2018.
\newblock \doi{10.32614/RJ-2018-037}.

\bibitem[Baumer et~al.(2014)Baumer, {\c C}etinkaya-Rundel, Bray, Loi, and
  Horton]{Baumer:2014wi}
Benjamin~S. Baumer, Mine {\c C}etinkaya-Rundel, Andrew Bray, Linda Loi, and
  Nicholas~J. Horton.
\newblock R {M}arkdown: Integrating a reproducible analysis tool into
  introductory statistics.
\newblock \emph{Technology Innovations in Statistics Education}, 8\penalty0
  (1), 2014.
\newblock URL \url{https://escholarship.org/uc/item/90b2f5xh}.

\bibitem[Behnel et~al.(2011)Behnel, Bradshaw, Citro, Dalcin, Seljebotn, and
  Smith]{behnel2010cython}
S.~Behnel, R.~Bradshaw, C.~Citro, L.~Dalcin, D.S. Seljebotn, and K.~Smith.
\newblock Cython: The best of both worlds.
\newblock \emph{Computing in Science Engineering}, 13\penalty0 (2):\penalty0 31
  --39, 2011.
\newblock ISSN 1521-9615.
\newblock \doi{10.1109/MCSE.2010.118}.

\bibitem[Beller et~al.(2014)Beller, Bacchelli, Zaidman, and
  Juergens]{Beller:2014}
Moritz Beller, Alberto Bacchelli, Andy Zaidman, and Elmar Juergens.
\newblock Modern code reviews in open-source projects: Which problems do they
  fix?
\newblock In \emph{{Proceedings of the 11th Working Conference on Mining
  Software Repositories}}, pages 202--211, 2014.
\newblock \doi{10.1145/2597073.2597082}.

\bibitem[Bentley(1975)]{Bentley_1975}
Jon~Louis Bentley.
\newblock Multidimensional binary search trees used for associative searching.
\newblock \emph{Communications of the ACM}, 18\penalty0 (9):\penalty0 509--517,
  1975.
\newblock \doi{10.1145/361002.361007}.

\bibitem[Bion et~al.(2018)Bion, Chang, and Goodman]{Bion:2018fn}
Ricardo Bion, Robert Chang, and Jason Goodman.
\newblock How {R} helps {A}irbnb make the most of its data.
\newblock \emph{The American Statistician}, 72\penalty0 (1):\penalty0 46--52,
  2018.
\newblock \doi{10.1080/00031305.2017.1392362}.

\bibitem[Bissi et~al.(2016)Bissi, Neto, and Emer]{Bissi_2016}
Wilson Bissi, Adolfo Gustavo Serra~Seca Neto, and Maria Claudia
  Figueiredo~Pereira Emer.
\newblock The effects of test driven development on internal quality, external
  quality and productivity: A systematic review.
\newblock \emph{Information and Software Technology}, 74:\penalty0 45--54,
  2016.
\newblock \doi{10.1016/j.infsof.2016.02.004}.

\bibitem[Breiman(2001)]{Breiman:2001}
Leo Breiman.
\newblock Random forests.
\newblock \emph{Machine Learning}, 45\penalty0 (1):\penalty0 5--32, 2001.
\newblock \doi{10.1023/A:1010933404324}.

\bibitem[Bryan(2018)]{Bryan:2018kq}
Jennifer Bryan.
\newblock Excuse me, do you have a moment to talk about version control?
\newblock \emph{The American Statistician}, 72\penalty0 (1):\penalty0 20--27,
  2018.
\newblock \doi{10.1080/00031305.2017.1399928}.

\bibitem[\c{C}etinkaya{-}Rundel and Rundel(2018)]{CetinkayaRundel:2018}
Mine \c{C}etinkaya{-}Rundel and Colin Rundel.
\newblock Infrastructure and tools for teaching computing throughout the
  statistical curriculum.
\newblock \emph{The American Statistician}, 72\penalty0 (1):\penalty0 58--65,
  2018.
\newblock \doi{10.1080/00031305.2017.1397549}.

\bibitem[{CMU Career \& Professional Development Center}(2018)]{CPDC:2018}
{CMU Career \& Professional Development Center}.
\newblock First destination outcomes: {D}ietrich {C}ollege {S}tatistics \&
  {D}ata {S}cience, bachelor's, 2018.
\newblock URL
  \url{https://www.cmu.edu/career/documents/2018_one_pagers/dc/Bachelors\%20Stats.pdf}.

\bibitem[Eddelbuettel and Francois(2011)]{Eddelbuettel:2011}
Dirk Eddelbuettel and Romain Francois.
\newblock Rcpp: Seamless {R} and {C++} integration.
\newblock \emph{Journal of Statistical Software}, 40\penalty0 (8), 2011.
\newblock \doi{10.18637/jss.v040.i08}.

\bibitem[Eklund et~al.(2016)Eklund, Nichols, and Knutsson]{Eklund:2016}
Anders Eklund, Thomas~E Nichols, and Hans Knutsson.
\newblock Cluster failure: Why f{MRI} inferences for spatial extent have
  inflated false-positive rates.
\newblock \emph{Proceedings of the National Academy of Sciences}, 113\penalty0
  (28):\penalty0 7900--7905, 2016.
\newblock \doi{10.1073/pnas.1602413113}.

\bibitem[Fiksel et~al.(2019)Fiksel, Jager, Hardin, and Taub]{Fiksel:2019}
Jacob Fiksel, Leah~R Jager, Johanna~S Hardin, and Margaret~A Taub.
\newblock Using {G}it{H}ub {C}lassroom to teach statistics.
\newblock \emph{Journal of Statistics Education}, 27\penalty0 (2):\penalty0
  110--119, 2019.
\newblock \doi{10.1080/10691898.2019.1617089}.

\bibitem[Freeman et~al.(2014)Freeman, Eddy, McDonough, Smith, Okoroafor, Jordt,
  and Wenderoth]{Freeman_2014}
S.~Freeman, S.~L. Eddy, M.~McDonough, M.~K. Smith, N.~Okoroafor, H.~Jordt, and
  M.~P. Wenderoth.
\newblock Active learning increases student performance in science,
  engineering, and mathematics.
\newblock \emph{Proceedings of the National Academy of Sciences}, 111\penalty0
  (23):\penalty0 8410--8415, May 2014.
\newblock \doi{10.1073/pnas.1319030111}.

\bibitem[Gray and Moore(2003)]{Gray_2003}
Alexander~G. Gray and Andrew~W. Moore.
\newblock Nonparametric density estimation: Toward computational tractability.
\newblock In \emph{Proceedings of the 2003 SIAM International Conference on
  Data Mining}, pages 203--211, 2003.
\newblock \doi{10.1137/1.9781611972733.19}.

\bibitem[Greenhouse and Seltman(2018)]{Greenhouse:2018}
Joel~B Greenhouse and Howard~J Seltman.
\newblock On teaching statistical practice: From novice to expert.
\newblock \emph{The American Statistician}, 72\penalty0 (2):\penalty0 147--154,
  2018.
\newblock \doi{10.1080/00031305.2016.1270230}.

\bibitem[Johnson(2013)]{Johnson_2013}
Nicholas~A Johnson.
\newblock A dynamic programming algorithm for the fused lasso and
  {$L_0$}-segmentation.
\newblock \emph{Journal of Computational and Graphical Statistics}, 22\penalty0
  (2):\penalty0 246--260, 2013.
\newblock \doi{10.1080/10618600.2012.681238}.

\bibitem[Jordan(2013)]{Jordan_2013}
Michael~I Jordan.
\newblock On statistics, computation and scalability.
\newblock \emph{Bernoulli}, 19\penalty0 (4):\penalty0 1378--1390, 2013.
\newblock \doi{10.3150/12-BEJSP17}.

\bibitem[Kulik et~al.(1990)Kulik, Kulik, and Bangert-Drowns]{Kulik:1990zp}
Chen-Lin~C Kulik, James~A Kulik, and Robert~L Bangert-Drowns.
\newblock Effectiveness of mastery learning programs: A meta-analysis.
\newblock \emph{Review of Educational Research}, 60\penalty0 (2):\penalty0
  265--299, 1990.
\newblock \doi{10.3102/00346543060002265}.

\bibitem[Leisch(2002)]{Leisch:2002}
Friedrich Leisch.
\newblock Sweave: Dynamic generation of statistical reports using literate data
  analysis.
\newblock In Wolfgang H{\"a}rdle and Bernd R{\"o}nz, editors, \emph{Compstat
  2002 --- Proceedings in Computational Statistics}, pages 575--580. Physica
  Verlag, Heidelberg, 2002.
\newblock ISBN 3-7908-1517-9.

\bibitem[Li-chun Wang(2003)]{Wang:2003}
Avery Li-chun Wang.
\newblock An industrial-strength audio search algorithm.
\newblock In \emph{Proceedings of the 4th International Conference on Music
  Information Retrieval}, 2003.

\bibitem[Liu et~al.(2012)Liu, Ting, and Zhou]{Liu:2012}
Fei~Tony Liu, Kai~Ming Ting, and Zhi-Hua Zhou.
\newblock Isolation-based anomaly detection.
\newblock \emph{ACM Transactions on Knowledge Discovery from Data}, 6\penalty0
  (1):\penalty0 3:1--3:39, March 2012.
\newblock \doi{10.1145/2133360.2133363}.

\bibitem[{Mäntylä} and {Lassenius}(2009)]{Mantyla:2009}
M.~V. {Mäntylä} and C.~{Lassenius}.
\newblock What types of defects are really discovered in code reviews?
\newblock \emph{IEEE Transactions on Software Engineering}, 35\penalty0
  (3):\penalty0 430--448, May 2009.
\newblock ISSN 2326-3881.
\newblock \doi{10.1109/TSE.2008.71}.

\bibitem[{National Academies of Sciences, Engineering, and
  Medicine}(2018)]{NAS:2018}
{National Academies of Sciences, Engineering, and Medicine}.
\newblock \emph{Data Science for Undergraduates: Opportunities and Options}.
\newblock The National Academies Press, Washington, DC, 2018.
\newblock \doi{10.17226/25104}.

\bibitem[Nolan and Temple~Lang(2009)]{Nolan:2009}
Deborah Nolan and Duncan Temple~Lang.
\newblock Integrating computing into the statistics curricula, 2009.
\newblock URL \url{https://www.stat.berkeley.edu/~statcur/}.

\bibitem[Nolan and Temple~Lang(2010)]{Nolan:2010}
Deborah Nolan and Duncan Temple~Lang.
\newblock Computing in the statistics curricula.
\newblock \emph{The American Statistician}, 64\penalty0 (2):\penalty0 97--107,
  2010.
\newblock \doi{10.1198/tast.2010.09132}.

\bibitem[Rigby and Bird(2013)]{Rigby:2013}
Peter~C Rigby and Christian Bird.
\newblock Convergent contemporary software peer review practices.
\newblock In \emph{Proceedings of the 9th Joint Meeting on Foundations of
  Software Engineering}, pages 202--212, 2013.
\newblock \doi{10.1145/2491411.2491444}.

\bibitem[Rossini(2001)]{Rossini:2001}
A~J Rossini.
\newblock Literate statistical practice.
\newblock In K~Hornik and F~Leisch, editors, \emph{Proceedings of the 2nd
  InternationalWorkshop on Distributed Statistical Computing}, 2001.

\bibitem[Sadowski et~al.(2018)Sadowski, S\"{o}derberg, Church, Sipko, and
  Bacchelli]{Sadowski:2018}
Caitlin Sadowski, Emma S\"{o}derberg, Luke Church, Michal Sipko, and Alberto
  Bacchelli.
\newblock Modern code review: A case study at {G}oogle.
\newblock In \emph{Proceedings of the 40th International Conference on Software
  Engineering: Software Engineering in Practice}, pages 181--190, 2018.
\newblock \doi{10.1145/3183519.3183525}.

\bibitem[Wayne(2016)]{Wayne:2016}
Kevin Wayne.
\newblock Autocomplete-me.
\newblock In \emph{SIGCSE Nifty Assignments}, 2016.
\newblock URL \url{http://nifty.stanford.edu/2016/wayne-autocomplete-me/}.

\bibitem[Wickham(2011)]{Wickham:2011}
Hadley Wickham.
\newblock testthat: Get started with testing.
\newblock \emph{The R Journal}, 3\penalty0 (1):\penalty0 5--10, 2011.
\newblock URL
  \url{https://journal.r-project.org/archive/2011-1/RJournal_2011-1_Wickham.pdf}.

\bibitem[Wickham(2014)]{Wickham_2014}
Hadley Wickham.
\newblock Tidy data.
\newblock \emph{Journal of Statistical Software}, 59\penalty0 (10), 2014.
\newblock \doi{10.18637/jss.v059.i10}.

\bibitem[Wickham et~al.(2019)Wickham, Averick, Bryan, Chang, McGowan, Fran{\c
  c}ois, Grolemund, Hayes, Henry, Hester, Kuhn, Pedersen, Miller, Bache,
  Müller, Ooms, Robinson, Seidel, Spinu, Takahashi, Vaughan, Wilke, Woo, and
  Yutani]{Wickham_2019}
Hadley Wickham, Mara Averick, Jennifer Bryan, Winston Chang, Lucy~D'Agostino
  McGowan, Romain Fran{\c c}ois, Garrett Grolemund, Alex Hayes, Lionel Henry,
  Jim Hester, Max Kuhn, Thomas~Lin Pedersen, Evan Miller, Stephan~Milton Bache,
  Kirill Müller, Jeroen Ooms, David Robinson, Dana~Paige Seidel, Vitalie
  Spinu, Kohske Takahashi, Davis Vaughan, Claus Wilke, Kara Woo, and Hiroaki
  Yutani.
\newblock Welcome to the tidyverse.
\newblock \emph{Journal of Open Source Software}, 4\penalty0 (43):\penalty0
  1686, Nov 2019.
\newblock \doi{10.21105/joss.01686}.

\bibitem[{Williams} et~al.(2003){Williams}, {Maximilien}, and
  {Vouk}]{Williams_2003}
L.~{Williams}, E.~M. {Maximilien}, and M.~{Vouk}.
\newblock Test-driven development as a defect-reduction practice.
\newblock In \emph{14th International Symposium on Software Reliability
  Engineering}, pages 34--45, Nov 2003.
\newblock \doi{10.1109/ISSRE.2003.1251029}.

\bibitem[Xie(2015)]{Xie:2015}
Yihui Xie.
\newblock \emph{Dynamic Documents with {R} and knitr}.
\newblock Chapman and Hall/CRC, Boca Raton, Florida, 2nd edition, 2015.
\newblock URL \url{https://yihui.name/knitr/}.
\newblock ISBN 978-1498716963.

\bibitem[Xie et~al.(2018)Xie, Allaire, and Grolemund]{Xie:2018}
Yihui Xie, J.J. Allaire, and Garrett Grolemund.
\newblock \emph{R Markdown: The Definitive Guide}.
\newblock Chapman and Hall/CRC, Boca Raton, Florida, 2018.
\newblock URL \url{https://bookdown.org/yihui/rmarkdown}.
\newblock ISBN 9781138359338.

\end{thebibliography}

\end{document}